\documentclass[twoside]{report}
\usepackage{iwsm}
\usepackage{graphicx}
\usepackage{amsmath, amssymb}
\usepackage{booktabs}

\begin{document}


\title{Bayesian shared-parameter models for analysing  sardine fishing in the Mediterranean Sea}
\titlerunning{Bayesian shared-parameter models}

\author{Gabriel Calvo\inst{1}, Carmen Armero\inst{1}, Maria Grazia Pennino\inst{2}, Luigi Spezia\inst{3}}
\authorrunning{Calvo et al.}    

\institute{Universitat de Val\`encia, Spain
\and Instituto Espa\~nol de Oceanograf\'ia, Spain
\and Biomathematics \& Statistics Scotland, Aberdeen, UK}

\email{gabriel.calvo@uv.es}

\abstract{European sardine is  experiencing an overfishing around the world. The dynamics of the industrial and artisanal fishing in the Mediterranean Sea   from 1970 to 2014 by country was assessed by means of     Bayesian joint longitudinal modelling that uses the random effects to generate an association structure between both longitudinal measures. Model selection was based on Bayes factors   approximated through the harmonic mean.}

\keywords{Joint modelling; Longitudinal data; Model comparison.}

\maketitle



\section{Introduction}

 European sardine (\textit{Sardina pilchardus}) is one of the most commercial species showing high over-exploitation rates over the last years in the Mediterranean Sea. Mediterranean fisheries are highly diverse and geographically varied due not only to the existence of different marine environments, but also because of different socio-economic situations and fisheries status.

We consider  data of European sardine landings from 1970 to 2014 from both the artisanal and the industrial fisheries which are defined in terms of  small-scale and large-scale commercial fisheries, respectively. Data are recorded by  country (Albania,
Algeria, Bosnia and Herzegovina, Croatia, France, Greece, Italy, Montenegro, Morocco,
Slovenia, Spain and Turkey) and come from \textsl{Sea Around Us} (www.seaaroundus.org), a research initiative at the University of British Columbia.

Top plot of the next figure shows the dynamics of the industrial amount of fish caught, in logarithmic
scale, from 1970 to 2014 in all the Mediterranean countries included in the study. Bottom plot presents the dynamics of artisanal fishing. \vspace*{-0cm}

	

\begin{center}
\includegraphics[width=80mm]{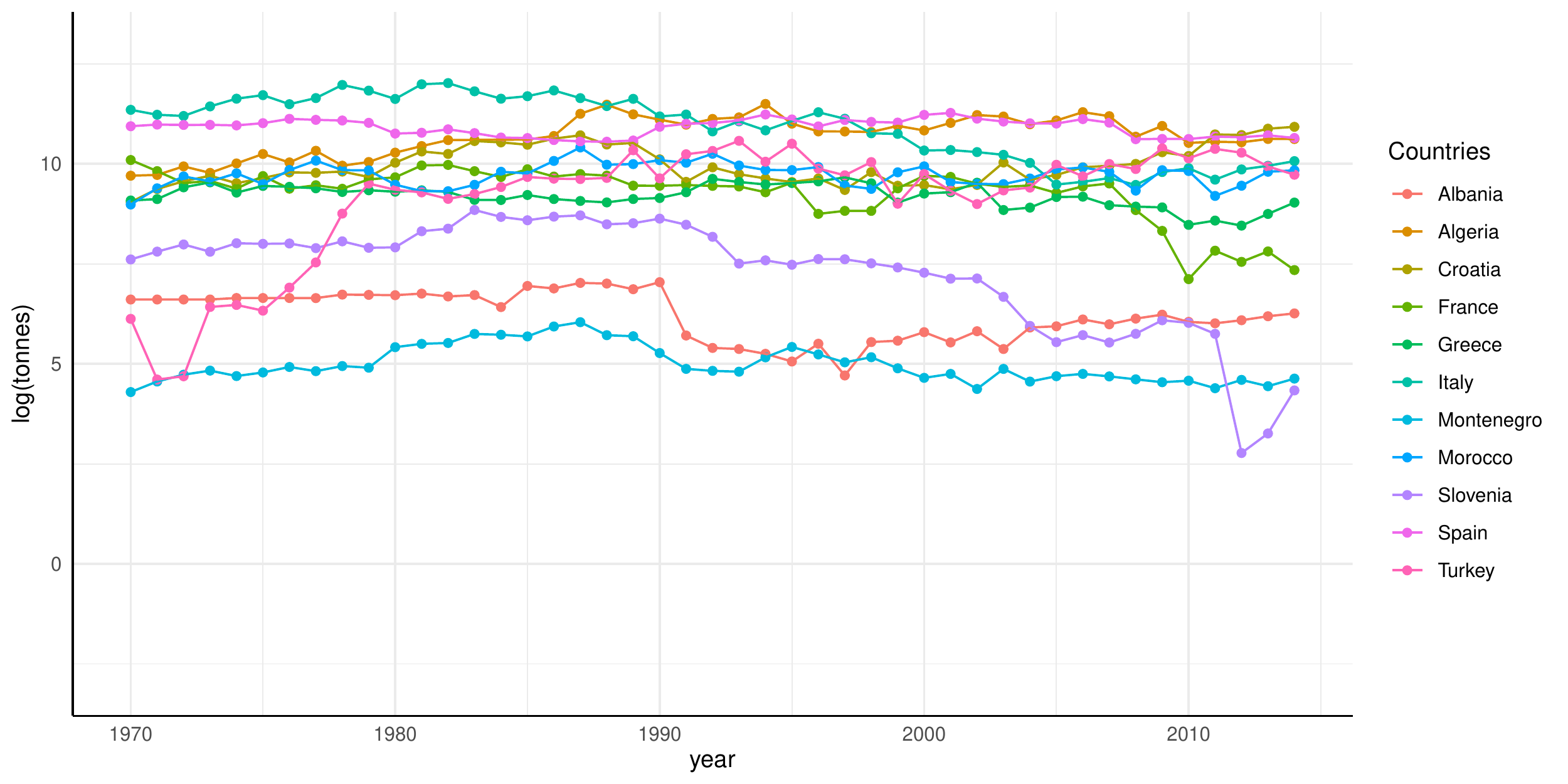}

\includegraphics[width=80mm]{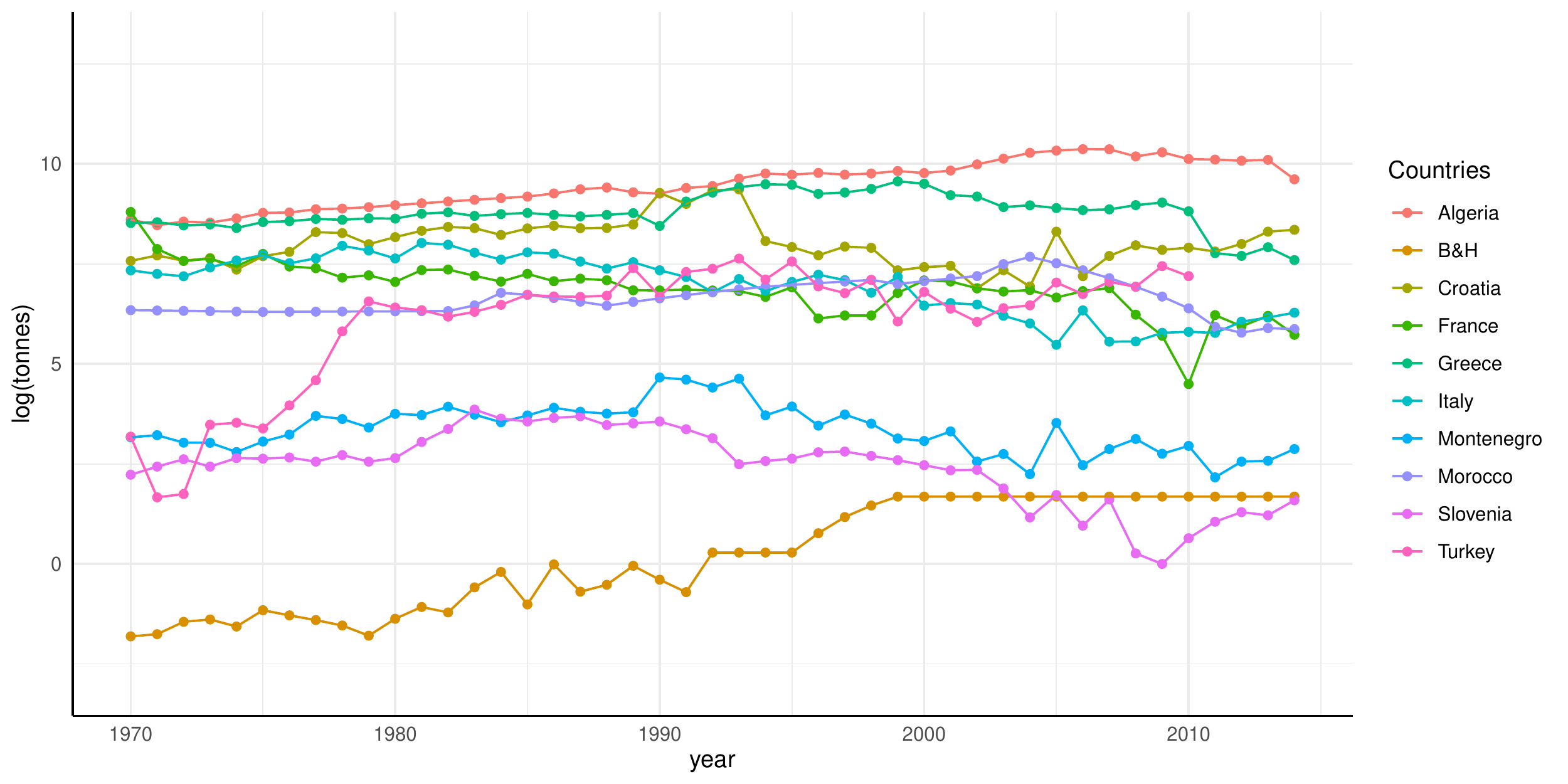}
\end{center}

\section{Bayesian joint modelling}
Let   $Y_{it}^{(I)}$ and $Y_{it}^{(A)}$ be the amount of sardine caught by industrial and artisanal methods in the country $i$  during year $t$, respectively. Calendar time is the time scale   and    $t=0$ is the first year of the study (i.e. $1970$).

We  assume  a Bayesian shared-parameter approach to jointly model both processes that uses the random effects to generate   an association structure  between both longitudinal measures. The joint distribution of the longitudinal fishing vectors, $\boldsymbol{Y}_{i}^{(I)}=(Y_{i0}^{(I)},\dots, Y_{iT}^{(I)} )$ and $\boldsymbol{Y}_{i}^{(A)}=(Y_{i0}^{(A)},\dots, Y_{iT}^{(A)} )$, parameters and hyperparameters $\boldsymbol{\theta}$, and random effects $\boldsymbol{b}_i$ for country $i$  is:
\begin{equation*}
\begin{split}
f(\boldsymbol{y}_{i}^{(I)}, \boldsymbol{y}_{i}^{(A)}, \boldsymbol{\theta}, \boldsymbol{b}_i)&= f(\boldsymbol{y}_{i}^{(I)}, \boldsymbol{y}_{i}^{(A)}| \boldsymbol{\theta}, \boldsymbol{b}_i)f(\boldsymbol{b}_i|\boldsymbol{\theta})\pi(\boldsymbol{\theta}) \\
&= f(\boldsymbol{y}_{i}^{(I)}| \boldsymbol{\theta}, \boldsymbol{b}_i^{(I)})
f(\boldsymbol{y}_{i}^{(A)}| \boldsymbol{\theta}, \boldsymbol{b}_i^{(A)})
f(\boldsymbol{b}_i|\boldsymbol{\theta})\pi(\boldsymbol{\theta}).
\end{split}
\end{equation*}
We propose two specific models for  $f(\boldsymbol{y}_{i}^{(I)}| \boldsymbol{\theta}, \boldsymbol{b}_i^{(I)})$ and
$f(\boldsymbol{y}_{i}^{(A)}| \boldsymbol{\theta}, \boldsymbol{b}_i^{(A)})$  within the framework of  mixed linear models.  The random effects vector for country $i$ can be divided in two subvectors, $\boldsymbol{b}_i = (\boldsymbol{b}_i^{(I)}, \boldsymbol{b}_i^{(A)}) $ corresponding to industrial and artisanal fishing, respectively. In addition, we impose a structure of association   between the random effects associated to the industrial  and artisanal fishing,  $f(b_{0i}^{(I)},b_{0i}^{(A)}|\Sigma_0)=\text{N}(0,\Sigma_0) $ and $f(b_{1i}^{(I)},b_{1i}^{(A)}|\Sigma_1)=\text{N}(0,\Sigma_1) $, with variance - covariance matrices given by
\begin{equation*}
	\Sigma_0 = \begin{pmatrix}
		{\sigma_0^{(I)}}^2 & \rho_0 \sigma_0^{(I)} \sigma_0^{(A)}\\
		\rho_0 \sigma_0^{(I)} \sigma_0^{(A)} & {\sigma_0^{(A)}}^2
	\end{pmatrix},\ \ \ \Sigma_1 = \begin{pmatrix}
		{\sigma_1^{(I)}}^2 & \rho_1 \sigma_1^{(I)} \sigma_1^{(A)}\\
		\rho_1 \sigma_1^{(I)} \sigma_1^{(A)} & {\sigma_1^{(A)}}^2
	\end{pmatrix}.
\end{equation*}
Both models can be expressed in terms of a conditional normal density of the subsequent longitudinal model given parameters, hyperparameters and random effects, 
$f(\boldsymbol{y}_{i}^{(I)}| \boldsymbol{\theta}, \boldsymbol{b}_{i}) = $ $  \text{N}(\boldsymbol{\mu}_{i}^{(I)},\sigma^2I),\,\,\,
f(\boldsymbol{y}_{i}^{(A)}| \boldsymbol{\theta}, \boldsymbol{b}_{i})=  \text{N}(\boldsymbol{\mu}_{i}^{(A)},  \sigma^2I). $

The two models differ in the conditional mean by the inclusion in one of them of an autoregressive  term  as it can be seen in Table \ref{calvo:tab:models}.
\begin{table}[h]
	\centering
	\caption{Conditional mean  of the amount of sardine    caught by industrial and artisanal methods in the country $i$  during year $t$   specified by each of the proposed   models. }
	\label{calvo:tab:models}
	\medskip
	\small
	\begin{tabular}{cll}
		\noalign{\hrule height 1pt}
		Model   & $ \mu_{it}^{(I)}$ & $\mu_{it}^{(A)}$\\
\noalign{\hrule height 1pt}
   $M1$ & $\beta_0^{(I)} + b_{0i}^{(I)} + b_{1i}^{(I)} t$ & $\beta_0^{(A)} + b_{0i}^{(A)} + b_{1i}^{(A)} t$\\
   $M2$ & $\beta_0^{(I)} + b_{0i}^{(I)} + b_{1i}^{(I)} t + \rho^{(I)} w_{i,t-1}^{(I)}$ & $\beta_0^{(A)} + b_{0i}^{(A)} + b_{1i}^{(A)} t + \rho^{(A)} w_{i,t-1}^{(A)}$\\
		\noalign{\hrule height 1pt}
	\end{tabular}
\normalsize
\end{table}

  Coefficients $\beta_0^{(I)}$ and $\beta_0^{(A)}$ are the regression  industrial and artisanal fishing intercept,  respectively. The autoregressive term in model $M2$   for industrial and artisanal fishing in country $i$ is  $w_{i,t-1}^{(I)} = y_{i,t-1}^{(I)} - (\beta_0^{(I)} + b_{0i}^{(I)} + b_{1i}^{(I)} (t-1))$ and $w_{i,t-1}^{(A)} = y_{i,t-1}^{(A)} - (\beta_0^{(A)} + b_{0i}^{(A)} + b_{1i}^{(A)} (t-1))$  (Weiss, 2005). 

  To fully specify the Bayesian model we elicit a prior distribution   for all  the uncertainties in the model.  We assume   a noninformative prior scenario with prior independence: normal distributions  for the regression coefficients and uniform distributions for all standard deviation  parameters. The prior   for the autoregressive parameters is  $\text{U}(-1,1)$ to induce the stationarity of $w_{it}^{(I)}$ and $w_{it}^{(A)}$, as well as for the correlation parameters $\rho_0$ and $\rho_1$.\vspace*{-0.1cm}

   \section{Posterior inferences}
The posterior distribution for both models was approximated via JAGS software (Plummer, 2003) through Markov chain Monte Carlo simulation. Table \ref{calvo:tab:posterior_results} summarizes the approximate posterior distribution for the models of our study.

Results indicate that the random effects  associated with each country play an important role in every model. Although the deviation of the random trends $\sigma_1^{(*)}$ has a small value, little variations on the trend produce big changes over time. Since the response variable is the logarithm of the tonnes, the random effects on the trend associated with each country play an important role in these models. On the other hand,   the inclusion of the  autoregressive term seems to absorb  a large part of the variability explained by the rest of the random effects and consequently, random effects become less relevant.\vspace*{-0.2cm}

\begin{table}[h]
	\centering
	\caption{Summary of the approximate sample from the posterior distribution for models $M1$ and $M2$. }
	\label{calvo:tab:posterior_results}
	\medskip
	\small
	\begin{tabular}{crrrrr}
		\noalign{\hrule height 1pt}
		\multicolumn{1}{c}{}                & \multicolumn{2}{c}{$M1$}   & &\multicolumn{2}{c}{$M2$}\\ \cline{2-3} \cline{5-6}
		\multicolumn{1}{c}{}      & \multicolumn{1}{c}{mean}     & \multicolumn{1}{c}{sd} & & \multicolumn{1}{c}{mean}     & \multicolumn{1}{c}{sd}             \\ \noalign{\hrule height 1pt}
		\multicolumn{1}{c}{$\beta_0^{(I)}$} & \multicolumn{1}{r}{$8.731$}  & $0.875$ & & $8.243$ & $0.793$                                \\
		\multicolumn{1}{c}{$\beta_0^{(A)}$} & \multicolumn{1}{r}{$5.651$}  & $1.208$  & & $5.707$ & $1.155$     \\		
		\multicolumn{1}{c}{$\rho_0$}      & \multicolumn{1}{r}{$0.673$}  & $0.258$  & & $0.695$ & $0.259$        \\
		\multicolumn{1}{c}{$\rho_1$}      & \multicolumn{1}{r}{$0.900$}  & $0.097$  & & $0.763$ & $0.253$         \\
		\multicolumn{1}{c}{$\rho^{(I)}$} & -  & -    & & $0.806$ & $0.042$    \\
		\multicolumn{1}{c}{$\rho^{(A)}$} & -  &-    & & $0.916$ & $0.035$               \\
		\multicolumn{1}{c}{$\sigma_{0}^{(I)}$}      & $2.648$  & $0.786$  & & $2.526$ & $0.745$         \\
		\multicolumn{1}{c}{$\sigma_{0}^{(A)}$}      & $3.823$  & $1.014$  & & $3.632$ & $1.046$     \\		
		\multicolumn{1}{c}{$\sigma_1^{(I)}$}    & \multicolumn{1}{r}{$0.051$} & $0.013$    & & $0.045$ & $0.013$                       \\
		\multicolumn{1}{c}{$\sigma_1^{(A)}$}    & \multicolumn{1}{r}{$0.052$}  & $0.012$ & & $0.037$ & $0.015$\\
		\multicolumn{1}{c}{$\sigma$}      & \multicolumn{1}{r}{$0.565$}  & $0.014$  & & $0.349$ & $0.008$                      \\
		\noalign{\hrule height 1pt}
	\end{tabular}
\normalsize
\end{table}\vspace*{-0.2cm}

As a first approach to  model comparison, we have computed the marginal likelihood for each model   by means of the harmonic mean (Newton and Raftery, 1994). The values obtained for  models  $M1$ and $M2$ in logarithmic scale are  $-817.74$ and $-523.27$,   respectively. The subsequent Bayes  factors   provide a decisive evidence in favour of the joint autoregressive model.
\vspace*{-0.5cm}

\acknowledgments{Calvo’s research was funded by  the ONCE Foundation and the Spanish Ministry of Education and Professional Training, grant FPU18/03101.  Spezia’s research was funded by the Scottish Government’s Rural and Environment Science and Analytical Services Division.} \vspace*{-0.6cm}



\references \vspace*{-0.1cm}
\begin{description}
\item[Armero, C., Forte, A., Perpiñán, H., Sanahuja, M. J. and Agustí, S.] (2018).
\vspace*{-0.1cm}

	 Bayesian  joint  modeling  for  assessing  the  progression  of  chronic kidney disease in children.
	 {\it Statistical Methods in Medical Research}, {\bf 27},
	 298\,--\,311.	
\item[Newton, M.A. and Raftery, A.E.] (1994).
     Approximate Bayesian inference with the weighted likelihood bootstrap.
	 {\it Journal of the Royal Statistical Society: Series B}, {\bf 56},
	 3\,--\,48.
\item[Plummer, M.] (2003).
     Jags: A program for analysis of Bayesian graphical models using Gibbs sampling.
     {\it Proceedings of the 3rd international workshop on distributed statistical computing}, {\bf 124},
     1\,--\,106.
\item[Weiss, R.E.] (2005).
     {\it Modeling longitudinal data}.
      Springer Science \& Business Media.
\end{description}

\end{document}